\documentstyle[openbib,12pt,fleqn]{article}
\newcommand{\bea}{\begin{eqnarray}}
\newcommand{\beq}{\begin{equation}}
\newcommand{\eea}{\end{eqnarray}}
\newcommand{\eeq}{\end{equation}}

\renewcommand{\baselinestretch}{2}

\begin{document}
\bibliographystyle{unsrt}

\setcounter{footnote}{0}

\begin{center}
\phantom{.}
{\Large \bf Breit-Wigner width for two interacting particles
in one-dimensional random potential \\}
{\small \sl   PH.JACQUOD~$^{(a)}$, D.L.SHEPELYANSKY~$^{(b,d,f)}$ and
 O.P.SUSHKOV~$^{(c,e,f)}$ \\}
 
{\small \it $^{(a)}$ Institut de Physique, Universit\'e de Neuch\^atel,\\
1, Rue A.L. Breguet, 2000 Neuch\^atel, Suisse \\
$^{(b)}$  Institute for Theoretical Physics,  University of California\\
Santa Barbara, CA 93106-4030\\
$^{(c)}$ 
Institute for Theoretical Atomic and Molecular Physics,\\
Harvard-Smithsonian Center for Astrophysics, MS14,\\
60 Garden Street, Cambridge, Massachusetts 02138}

\vspace{0.5truecm}

\vskip .3 truecm

\vspace{0.5truecm}
\end{center}
\small
{\bf Abstract:\/}
For two interacting particles (TIP) in one-dimensional
random potential the dependence of the Breit-Wigner
width $\Gamma$, the local density of states and the TIP localization length
on system parameters is determined analytically . The theoretical
predictions for $\Gamma$ are confirmed by numerical simulations.

\vskip .6 truecm


\newpage


Recently, the problem of two interacting particles (TIP) in a random
potential has attracted interest of different
groups  \cite{TIP,Imry,Pichard,Oppen,Borg1,Frahm3d}. It has been
shown that two repulsive/attracting particles can propagate together
on a distance $l_c$ much larger than one-particle localization length $l_1$
in absence of interaction. The first analytical studies \cite{TIP,Imry}
for TIP with on site interaction on a one-dimensional one channel lattice
gave the following estimate 
$l_c/l_1 \sim \Gamma \rho
\sim (U/V)^2 l_1$, where $U$ is strength of the interaction,
$V$ is intersite hopping matrix element, $\rho \sim l_1^2/V$ is density
of the two-particle states coupled by the interaction,
and $\Gamma \sim U^2/Vl_1$ is the interaction induced transition rate 
between these states.
The numerical investigations \cite{Pichard,Oppen} definitely 
confirmed existence of the strong enhancement of $l_c$
due to interaction. However, a direct verification of the above estimate
is quite difficult even for the modern computer facilities
due to the strong increase of required basis  with $l_1$.
Also the recent numerical results of von Oppen et. al. \cite{Oppen} 
and Weinmann and Pichard \cite{Pichard96} 
indicate in one-dimensional case almost linear growth of 
the enhancement factor for $l_c$ with $U$ instead of expected $U^2$.
Due to all these things it would be important to have a more rigorous
derivation of the factor $l_c/l_1$ for this
on a first glance quite simple problem, at least in a one-dimensional
case. To reach this aim we started from the computation of the rate
$\Gamma$ which also characterizes the spread width of the
Breit-Wigner distribution for eigenfunctions in the basis
of eigenstates of noninteracting particles\cite{Jaq,Fyod,Frahm1}. If the 
parameter dependence
of $\Gamma$ is known then the ratio $l_c/l_1$ can be determined
from the relation $l_c/l_1 \sim \Gamma \rho$ which have been checked
in models of superimposed band random matrices \cite{TIP,Jaq,Fyod,Frahm1}.
In the present work for calculation of $\Gamma$ we use the  technique
developed in \cite{Oleg} which allows to account all orders in the
interaction.

We consider one dimensional Hubbard model with Hamiltonian
\begin{equation}
\label{H}
H=-V\sum_{n \sigma}(a^{\dag}_{n+1 \sigma}a_{n \sigma} + a^{\dag}_{n \sigma}
a_{n+1 \sigma})+U\sum_n a_{n\uparrow}^{\dag}a_{n\downarrow}^{\dag}
a_{n\downarrow}a_{n\uparrow}
\end{equation}
Here $a^{\dag}_n$ is a creation operator of the particle at the site $n$,
$V$ is hopping matrix element, and $U$ is on site interaction.
We assume that particles are distinguishable and denote the type of
particle by spin $\sigma = \pm 1/2$.
Single particle eigenstate is plane wave 
$|p\rangle = {1\over{\sqrt{L}}}e^{ipn}$
with dispersion $\epsilon_p=-2V\cos p$, $-\pi \le p \le \pi$.
We set lattice spacing equal to unity. The size of the lattice is
denoted by $L$.

The Breit-Wigner width
can be found in the following way. Forward
scattering amplitude $f$ for particles with different spins is
given by series of diagrams presented at Fig.1. Solid line represents
a particle, and wavy line is matrix element of the interaction
$\langle p_3 p_4|\hat U|p_1 p_2 \rangle =
{U\over{L}}\delta_{p_1+p_2,p_3+p_4}$. Due to optical theorem width
of the state $|p_1 p_2 \rangle=|p_1\rangle |p_2 \rangle$ is related
to the forward scattering amplitude:
\begin{equation}
\label{opt}
\Gamma/2=-Im \ f.
\end{equation}
One can easily check the coefficient in this relation considering diagram
Fig. 1b which gives usual Fermi golden rule:
\begin{eqnarray}
\label{gold}
\Gamma \approx -2 \ Im f_{1b}&=&-2 \ Im \ \sum_{p_3 p_4}
{{\left|\langle p_3 p_4|\hat U|p_1 p_2 \rangle \right|^2}
\over{E-\epsilon_3 -\epsilon_4 +i0}}=\\
&=&2 \pi \sum_{p_3 p_4}
\left|\langle p_3 p_4|\hat U|p_1 p_2 \rangle \right|^2
\delta(E-\epsilon_3 -\epsilon_4).\nonumber
\end{eqnarray}
Here $E$ is energy of the initial state $E=\epsilon_1+\epsilon_2$.

Born term in the amplitude $f$ is given by Fig. 1a and equals $f_{1a}=U/L$. 
Calculation of the diagram Fig. 1b is also straightforward
\begin{eqnarray}
\label{box}
f_{1b}&=& \sum_{p_3 p_4}
{{\left|\langle p_3 p_4|\hat U|p_1 p_2 \rangle \right|^2}
\over{E-\epsilon_3 -\epsilon_4 +i0}}
={{U^2}\over{L^2}} \sum_{p_3}
{1\over{E+2V \cos p_3 + 2V \cos(p-p_3)}}=\nonumber\\
&=&{{U^2}\over{L^2}} \int_{-\pi}^{\pi}
{{L dp_3/2\pi}\over{\left[E+2V \cos p_3 + 2V \cos(p-p_3)\right]}}
={{U^2/L}\over{\sqrt{E^2-16V^2 \cos^2 p/2}}},
\end{eqnarray}
where $p=p_1+p_2=p_3+p_4$ is total quasi-momentum.
Higher orders in Fig. 1 correspond to simple iterations of the box Fig. 1b.
Therefore summation of the ladder is reduced to geometrical
progression and the result is
\begin{equation}
\label{ampl}
f(E,p)={{U/L}\over{1-U/\sqrt{E^2-16V^2 \cos^2 p/2}}}.
\end{equation}
The scattering amplitude depends only on total energy $-4V \le E \le 4V$
and total momentum $-\pi \le p \le \pi$. The branch of square root
should be chosen in such a way that $Im \ f \le 0$.

With amplitude (\ref{ampl}) one can easily calculate the Breit-Wigner width
using optical theorem (\ref{opt}). But we are interested in the average
width at given energy. So we have to average over momentum $p$.
Density of the two particle states is of the form
\begin{eqnarray}
\label{rpo2}
\rho(E,p)&=&\int_{-\pi}^{\pi}{{Ldp_1}\over{2\pi}}
\int_{-\pi}^{\pi}{{Ldp_2}\over{2\pi}}
\delta(p-p_1-p_2)\delta(E+2V \cos p_1 + 2V \cos p_2)=\nonumber \\
&=&{{L^2/(8\pi^2 V)}\over{\sqrt{\cos^2p/2 -E^2/16V^2}}}.
\end{eqnarray}
It is nonzero only if square root is real. After integration over
momenta we find
\begin{equation}
\label{ro2}
\rho(E)=\int_{-\pi}^{\pi}\rho(E,p){{dp}\over{2\pi}}
\approx {{L^2}\over{2\pi^2 V}}\left( \ln {{16 V}\over{|E|}}
+0.18 {{|E|}\over{4 V}}\right)
\end{equation}
The integral in (\ref{ro2}) can not be exactly expressed in terms
of elementary functions. Presented approximate formula is valid with 
accuracy better than 1\% in the interval  $-4 V \le E \le 4 V$. Now we can 
find the average Breit-Wigner width.
\begin{eqnarray}
\label{gav}
 \Gamma (E)  &=& -2 \ Im \ \int f(E,p) \rho(E,p)
{{dp}\over{2\pi}} \biggl/ \rho(E) \biggr.= \nonumber\\
&=&{{8 V u^2/L}\over{\left(\ln 4/\epsilon +0.18 \epsilon \right)
\sqrt{|u^2-{\epsilon}^2|(1+u^2-{\epsilon}^2)}}} \cdot F(Z).
\end{eqnarray}
Here $u =\left|U/4 V \right| $ and $\epsilon =\left|E/4 V \right|$ 
is interaction and energy  expressed in units of band width:
$0 \le \epsilon \le 1$. The function $F(Z)$ is defined by
\begin{eqnarray}
\label{F}
F(Z)&=&\left\{ \begin{array}{ll}
         \arctan Z, & \mbox{\ \ \ for u $\ge$ $\epsilon$} \\
         {1\over 2}\ln{{1+Z}\over{1-Z}}, & \mbox{\ \ \ for u $\le$ $\epsilon$}
        \end{array}
\right.\\
Z&=&\sqrt{{|u^2-{\epsilon}^2|(1-{\epsilon}^2)}
\over{{\epsilon}^2(1+u^2-{\epsilon}^2)}}.\nonumber
\end{eqnarray}
At small energy (${\epsilon}^2 \ll u^2, 1$)  formula (\ref{gav}) can be
substantially simplified
\begin{equation}
\label{gav1}
\Gamma \approx {{4 \pi V}\over{L}}\cdot {1\over{ \ln 4/\epsilon}}
\cdot {{u}\over{\sqrt{1+u^2}}},
\end{equation}
so that at small interaction ($\epsilon^2 \ll u^2 \ll 1$) it is linear in the interaction.
In other limit ($u^2 \ll {\epsilon}^2, \ (1-{\epsilon}^2)$) the width (\ref{gav}) is
quadratic in the interaction with logarithmic correction:
\begin{equation}
\label{gav2}
\Gamma \approx {{8 V}\over{L}}\cdot {1\over{(\ln 4/\epsilon +0.18 \epsilon)}}
\cdot {{u^2}\over{\epsilon}}\ln {{2 \epsilon \sqrt{1-{\epsilon}^2}}\over{u}}
\end{equation}
The value of $\Gamma$ in (\ref{gav1}) is significantly larger than in 
(\ref{gav2}) due to the growth of two-particle density of states (\ref{ro2})
near the center of the band.

If we now add to the Hamiltonian (\ref{H}) a single particle random potential
$H_{rand}=\sum w_n a^{\dag}_{n \sigma} a_{n \sigma}$
with a disorder homogeneously
distributed in the interval $-W \leq w_n \leq W$ , then one particle
eigenstates in infinite lattice become localized with localization length 
$l_1 \approx 24 (V/W)^2 \sqrt{1-\epsilon_1^2/4 V^2}$, where $\epsilon_1$ is 
one particle energy.
However as soon as $l_1 \gg 1$ 
the above calculation of the average
width remains valid. The reason for this is that $l_1 \gg 1$ is the
only condition which we need to formulate scattering problem and to
use conventional diagram technique.  
Distribution of $\Gamma$ depends on the relation between size of the
box $L$ and the localization length $l_1$. If $L \le l_1$ all 
values of $\Gamma$ are of the order of the average value given by (\ref{gav}).
For $L \gg l_1$ the average value is still given by (\ref{gav}).
However in this case $\Gamma$ vanishes for majority of the states . These
are the states in which particles are localized far from each other and 
practically do not interact. On other hand the width for 
the states with interparticle distance of the order $l_1$  
is approximately the same as for particles in a box of size
$L \approx l_1$ so that $\Gamma$ is given by
eqs.(\ref{gav}),(\ref{F}) with  $L$ replaced by $l_1$.
The two-particle localization length $l_c$ for such states
is determined by the relation $l_c/l_1 \sim \Gamma(E) \rho(E)$,
with $\Gamma$ calculated at $L \sim l_1$. 
This relation is valid if many unperturbed states are mixed by interaction 
\cite{TIP,Imry} so that $\Gamma(E) \rho(E) > 1$. In the opposite case
$\Gamma(E) \rho(E) \ll 1$ the above relation is not valid
\cite{Jaq,Fyod,Frahm1} and the interaction can be treated
in a perturbative way. In this regime  "Rabi oscillations" between
two quasi-degenerate levels play an important role \cite{Pichard96}.

Above we have considered distinguishable particles. The generalization to
identical particles is rather simple: the width $\Gamma$ vanishes if coordinate
wave function is antisymmetric, and it is doubled in comparison with
eqs.(\ref{gav}), (\ref{F}) if coordinate wave function is symmetric.

To check the above theoretical formula for the Breit-Wigner width $\Gamma$
we studied numerically  the model (\ref{H}) of two identical interacting
particles (symmetric coordinate wave function)
in  the disordered potential 
on a ring
of size $L$ which is
less or comparable with one-particle localization length 
$l_1 \approx 24 (V/W)^2$.  Using Lanczos technique (see for example
\cite{Dagotto}) we determined the local density of states
for symmetric configurations in the basis of noninteracting eigenstates:
\begin{equation}
\label{lds}
{\rho_{W}} (E-\epsilon_{m_1}-\epsilon_{m_2})={\sum_{\lambda}}
|\psi_{\lambda} (m_1,m_2)|^2 \delta(E-E_{\lambda})
\end{equation}
Here   $E_{\lambda}$ is the eigenenergy of TIP while $\epsilon_{m_{1,2}}$
are one-particle eigenenergies.  The  dependence of ${\rho}_{W}$
on $E$ is well described by the Breit-Wigner distribution
\begin{equation}
\label{BW}
{\rho_{W}} (E)={{\Gamma} \over {2\pi [E^2+\Gamma^2/4]}}
\end{equation}
an example of which is shown in Fig.2 . The comparison of
numerically obtained $\Gamma$ with theoretical prediction 
(\ref{gav}), (\ref{F}) 
in the regime $\Gamma(E) \rho(E) > 1$
is shown in Figs.3,4 for different energies as the function of interaction.
The theory gives good agreement with numerical results 
for $15 \leq L \leq 300$
and variation of scaled width $\Gamma L/V$ by more than 
2 orders of magnitude. For the states with the energy
close to the band center $(E \approx 0)$ (Fig.3) the dependence
of $\Gamma \rho$ on $U$ is almost linear for $U < V$ 
(see (\ref{ro2}), (\ref{gav1})).
Therefore, the TIP localization length $l_c$ according to the
relation $l_c/l_1=C \Gamma \rho \approx 2 C l_1 (U/V)/\pi$ also varies
linearly with $U$. Here, we took the values
of $\Gamma$ and $\rho$ at $L=l_1$
and introduced the numerical coefficient $C$ to take into account
the uncertainty of this choice. 
According to the numerical
results \cite{Oppen} at the center of the band $l_c/l_1 \approx 0.2 l_1 (U/V)$
which is in good agreement with the above theoretical
expression and gives $C \approx 1/4$.  

For energies away from the band 
center and small interaction $|U| \ll |E|$ the enhancement
factor according to (\ref{ro2}), (\ref{gav2}) is
$l_c/l_1 \approx l_1 U^2 \ln(2E/U)/(4{\pi}^2VE)$ where
we have used the above value of $C$. The dependence on $U$
is almost quadratic in agreement with the first 
estimate  \cite{TIP,Imry}. 
However, due to the logarithmic correction,
to observe clearly the $U^2$ behavior one should go to really small
$U$ values and since the condition $\Gamma \rho > 1$ should be also 
satisfied this can be reached only for quite large values of
$l_1$ or $L$. In this respect our numerical approach based on 
the measurement of $\Gamma$ is more efficient 
than the one used in \cite{Oppen}. It allows to see the behavior
$U^2 \ln U$ away from the band center in agreement with the
theory (\ref{gav}), (\ref{F}) (see insert in Fig.4).
At moderate $U/V > 0.3$ values in the presence of numerical fluctuations
the dependence of $\Gamma$ on $U$ is hardly distinguishable
from a linear one (see normal scale in Fig.4). In our opinion
this is the reason why the linear behavior in $U$ had been
attributed in \cite{Oppen} also to the states away from the band center.
As for the result of Ref. \cite{Pichard96} the system size there
was too small ($L=25$) and the main part of the data (Fig. 4
with $U/V < 0.4$) corresponds to the different regime $\Gamma \rho < 1$.
In this perturbative case the typical energy scale which
determines the change in level statistics is determined by Rabi  oscillation
frequency in a pair of quasi-degenerate states which
is proportional to $U$ \cite{Pichard96}. Also, one should keep in mind
that the results there are integrated over the whole energy band
including the center of the band where the dependence on $U$ is linear 
even for $\Gamma \rho > 1$.

Turning back to our numerical data (Fig. 4) 
we would like to mention that there is a significant difference
from the theory for negative $U < -1.$. Generally, we 
should expect such difference for $\vert U \vert \gg 1$
when the spectrum is composed from two separated energy
bands and the basis of plane waves used for computation
of width $\Gamma$ becomes inadequate. For example, 
in this regime the density of states is not  described by 
(\ref{ro2}). However, we cannot say
why this change goes in so asymmetric way for negative
and positive $U$ while for $\vert U \vert < 1$
the width $\Gamma$ is independent of sign $U$ in agreement 
with the theory. We would like to note that
such asymmetry for attraction and repulsion away from the band
center and relatively strong
interaction $U \approx V$ has been seen recently in \cite{Oppen}
for the ratio $l_c/l_1$. Also a change in the behavior of $\Gamma$
has been observed in \cite{Pichard96} for $U > V$.

In summary, taking diagrammatically
into account the effects of interaction we have
derived the analytical formula for the Breit-Wigner width $\Gamma$
which determines the enhancement factor $\l_c/l_1 \sim \Gamma \rho > 1$
for TIP in one-dimensional random potential. 
Our analytical and numerical approaches can be also used for
calculation of the TIP width in 2- and 3-dimensional 
disordered systems where according to Imry estimate \cite{Imry}
interaction between two quasi-particles can strongly affect transport
properties.

We acknowledge fruitful discussions with V.V.~Flambaum,
M.Yu.~Kuchiev and P.G.Silvestrov. We also thank the Centro Svizzero di Calcolo
Scientifico for allocation of CPU time on their NEC SX-3.
One of us (OPS) thanks Laboratoire de Physique Quantique,
Universit\'e Paul Sabatier for hospitality and financial support during the 
initial stage of this work.  This research was supported in part by the
National Science Foundation under Grant No. PHY94-07194,
NSF through a grant for the Institute for Theoretical Atomic and Molecular
Physics at Harvard University and the Smithsonian Astrophysical
Observatory and the Fonds National Suisse de la Recherche.

\vfill\eject

\renewcommand{\baselinestretch} {2}

\vfill\eject
{\bf {Figure captions}}
\vskip 20pt
\begin{description}
{
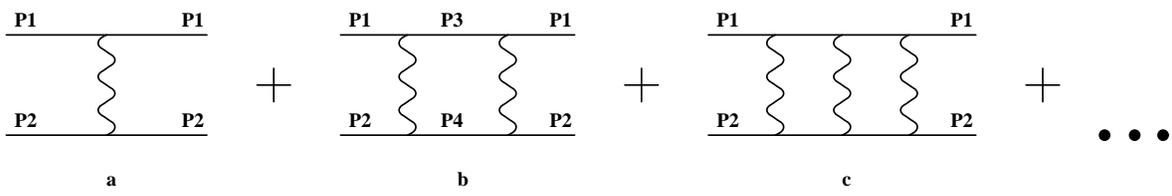
\item[Fig. 1:]  Diagrams for the forward scattering amplitude $f$
in (2) - (5).

\item[Fig. 2:]  Local spectral density $\rho_W (E)$
computed for the TIP eigenstates in the energy interval
$[-0.1,0.1]$ for the case $L=150$,
$U=1$, $V=1$, and $W=0.4$. The full line gives the best Breit-Wigner fit (13) with
$\Gamma = 0.0073$. The theoretical prediction is $\Gamma = 0.0072$.

\item[Fig. 3:] Scaled Breit-Wigner width $\Gamma L/V$ as a function of the
rescaled interaction $\frac{U}{4 V}$ computed in 
the energy interval $E/V \in [-0.1,0.1]$. The system size is $L=15$ 
($W/V=1.$, empty circles), $L=25$ ($W/V=1.$, empty squares), $L=40$ 
($W/V=0.6$, empty diamonds), $L=60$ ($W/V=0.5$, full circles), $L=80$ 
($W/V=0.5$, full squares), $L=100$ ($W/V=0.5$, full diamonds) $L=150$ 
($W/V=0.4$, full triangles up) and $L=200$ ($W/V=0.35$, full triangles down). 
The solid line gives the theoretical prediction (8), (9) multiplied by
2 to take symmetrization into account.

\item[Fig. 4:] Scaled Breit-Wigner width $\Gamma L/V$ as a function of the
rescaled absolute value of the interaction $\frac{|U|}{4 V}$ computed in 
the energy interval $E/V \in [1.,1.2]$. 
The system size is $L=15$ ($W/V=1.$, empty circles), 
$L=25$ ($W/V=1.$, empty squares and $W/V=0.5$, empty diamonds), $L=40$ 
($W/V=0.8$, empty triangles up and $W/V=0.5$, empty triangles down), 
$L=60$ ($W/V=0.5$, full circles), $L=80$ ($W/V=0.5$, full squares), $L=100$ 
($W/V=0.5$, full diamonds), $L=100$ 
($W/V=0.5$, negative U, crosses), $L=150$ ($W/V=0.4$, full triangles up), 
$L=200$ ($W/V=0.35$, full triangles down) and $L=300$ ($W/V=0.25$, full triangles left). 

}
\end{description}
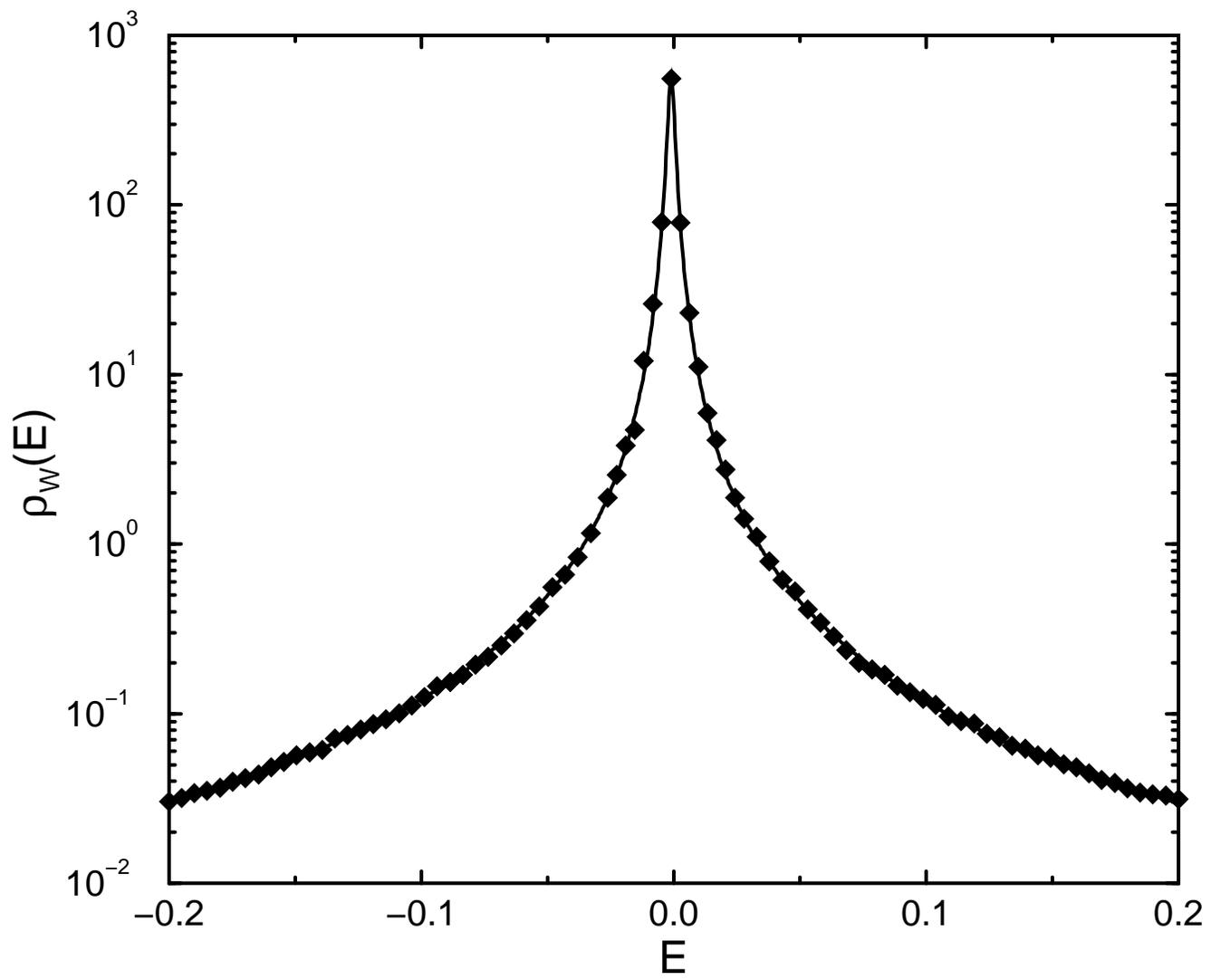
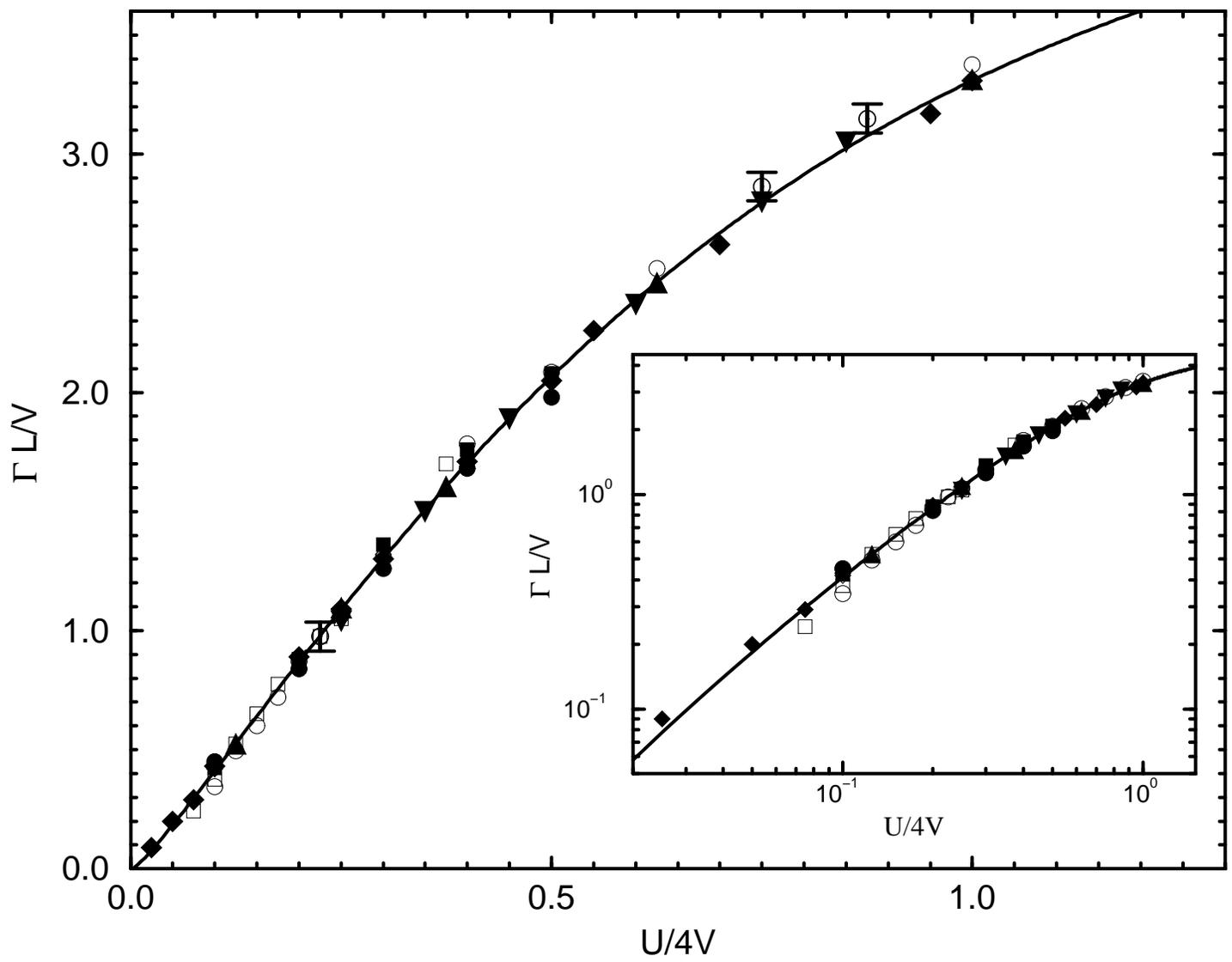
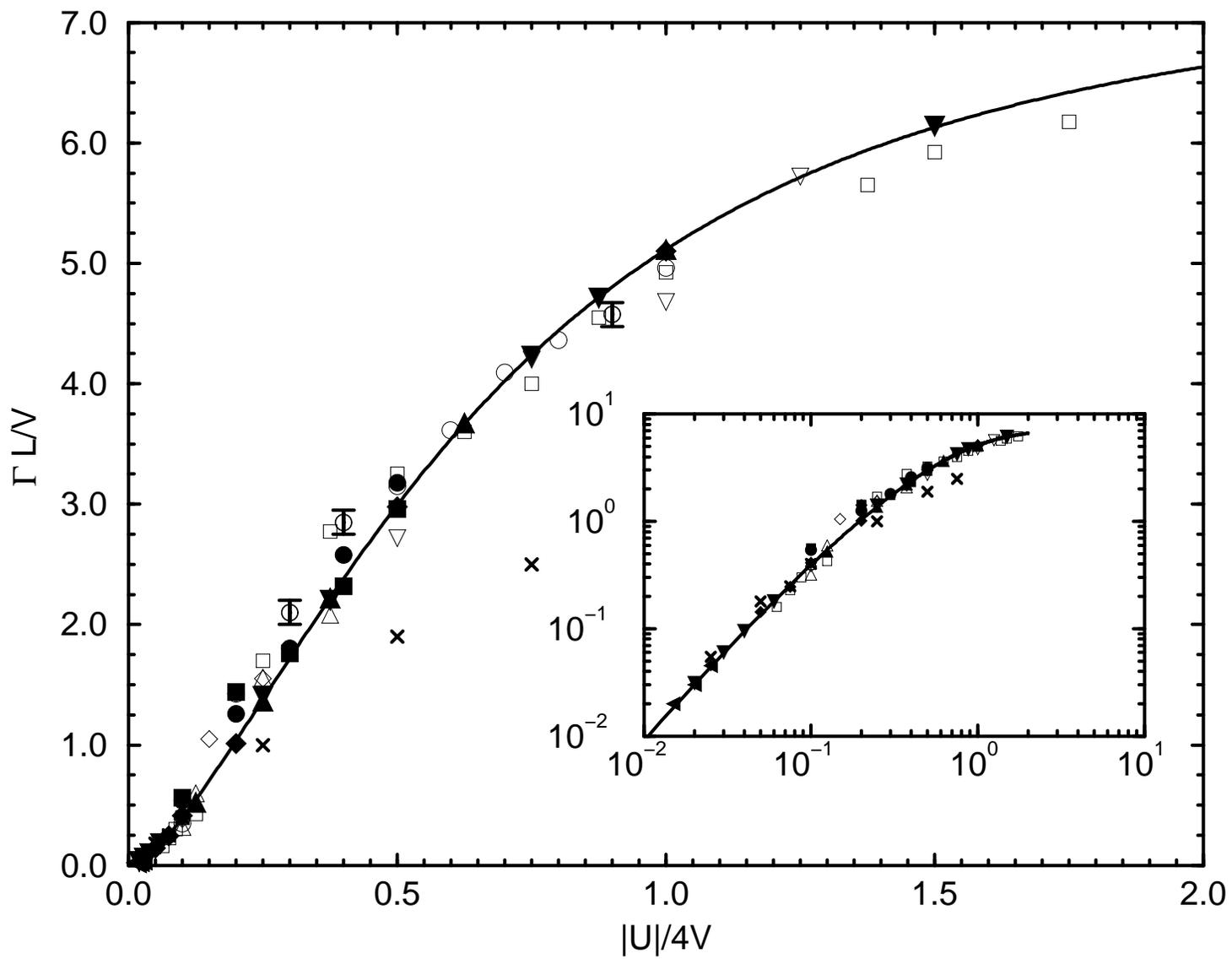
\end{document}